\documentclass[
    aps,12pt,final,notitlepage,
    oneside,onecolumn,nofootinbib,
    superscriptaddress,noshowpacs,
    nolongbibliography,centertags
]{revtex4-2}

\usepackage[english]{babel}

\usepackage[usenames,dvipsnames]{xcolor} 
\usepackage[
    colorlinks = true,
    urlcolor = MidnightBlue,
    linkcolor = Red,
    citecolor = NavyBlue
]{hyperref}

\usepackage{graphicx}
\graphicspath{{figures/}}

\usepackage{amsmath,amssymb}

\newcommand{\nthick}{\thickmuskip=0mu} 

\newcommand{\eqn}[1]{(\ref{#1})}
\newcommand{\fign}[1]{fig.~\ref{#1}}

\newcommand{\DOI}[2]{\href{https://doi.org/#1}{#2}}
\newcommand{\arXiv}[2]{arXiv:~\href{https://arXiv.org/abs/#1}{#1} [#2]}

\newcommand{\eq}[1]{(\ref{#1})}

\newcommand{{\yeas}}{{YEASA}}

\newcommand{\ethr}{\epsilon_{\text{thr.}}} 
\newcommand{\E}{E_0}
\newcommand{\dP}{d_p}
\newcommand{\dFe}{d_{\text{Fe}}}
\newcommand{\dExp}{d_{\text{exp}}}

\newcommand{\fs}{f_{\text{s}}} 
\newcommand{\fsh}{f_{\text{s}}} 
\newcommand{\rhos}{\rho_{\text{SD}}}
\newcommand{\rhosRT}{\rho_{\text{SD}}(r,\theta)}

\newcommand{\rhosSOOT}{\rho_{\text{SD}}(600,\theta)}
\newcommand{\rhosSOOV}{\rho_{\text{SD}}(600,0\degr)}
\newcommand{\bs}{b_{\text{s}}} 

\newcommand{\fmu}{f_{\text{MD}}}
\newcommand{\rhom}{\rho_{\text{MD}}}
\newcommand{\rhomRT}{\rho_{\text{MD}}(r,\theta)}
\newcommand{\rhomSOOT}{\rho_{\text{MD}}(600,\theta)}
\newcommand{\bmu}{b_{\text{MD}}}

\newcommand{\meancos}{\left<\cos\theta\right>}
\newcommand{\rM}{r_{\text{M}}}

\newcommand{\wmax}{w_{\text{max}}}

\newcommand{{\lnA}}{\langle\ln{A}\rangle} 
\newcommand{{\xmax}}{x_{\text{max}}}      
\newcommand{{\Xmp}}{x_{\text{max}}^p}
\newcommand{{\XmFe}}{x_{\text{max}}^{\text{Fe}}}
\newcommand{{\XmExp}}{x_{\text{max}}^{\text{exp}}}

\newcommand{\qgs}{{\sc qgsj}et01}
\newcommand{\qgsii}{{\sc qgsj}et-II.04}

\newcommand{\eposlhc}{{\sc epos-lhc}}
\newcommand{\sibyllold}{{\sc sibyll}-2.1}

\newcommand{\fluka}{{\sc fluka}2011}
\newcommand{\corsika}{{\sc corsika}}

\newcommand{{\usec}}{{~$\mu$c}}    
\newcommand{{\sqrm}}{{~m$^2$}}     
\newcommand{{\sqrkm}}{{~km$^2$}}   
\newcommand{\depth}{{~g/cm$^2$}}   
\newcommand{\dens}{{~g/cm$^3$}}    
\newcommand{\degr}{^{\circ}}       

\begin{document}

\title{Zenith-Angular Characteristics of Particles in EASs with $\E \simeq 10^{18}$~eV According to the Yakutsk Array Data}

\author{A.\,V.~Glushkov}
\email{glushkov@ikfia.ysn.ru}

\author{K.\,G.~Lebedev}
\affiliation{Yu.\,G.\,Shafer Institute of cosmophysical research and aeronomy of Siberian branch of the Russian Academy of Sciences \\
31 Lenin ave., Yakutsk, 677027, Russia}

\author{A.\,V.~Saburov}

\begin{abstract}
    Particle lateral distributions were investigated in cosmic ray air showers with energy $\E \simeq 10^{18}$~eV registered at the Yakutsk array with surface and underground scintillation detectors with $\simeq 1 \times \sec\theta$~GeV threshold during the period of continuous observations from 1986 to 2016. The analysis covers events with arrival direction zenith angles $\theta \le 60\degr$ within five intervals with step $\Delta\cos\theta = 0.1$. Experimental values were compared to simulation results obtained with the use of \corsika{} code within the framework of \qgs{} hadron interaction model. The whole dataset points at probable cosmic ray composition which is close to protons.
\end{abstract}

\maketitle

\section{Introduction}

Ultra-high energy cosmic rays (CRs) (with energy $\nthick\ge 10^{15}$~eV) have been actively studied worldwide for more than 50 years~\cite{Grieder(2010)}. Their mass composition is still haven't been measured precisely and without this knowledge it is difficult to understand the nature of hadronic interactions occurring at these energies and to identify the sources and origins of primary particles. The CR mass composition can be estimated by different air shower parameters ($d$) that are sensitive to it. At the Yakutsk array this is done via lateral distribution functions (LDFs) of electron, muon and Cherenkov component of extensive air showers (EAS) (see e.g. \cite{bib:2, bib:3, bib:4, bib:5, bib:6, bib:7}). The key to this problem is a simple relation, which follows from the principle of nucleonic superposition:

\begin{displaymath}
    \lnA = \frac{\dP - \dExp}{\dP - \dFe}
    \cdot
    \ln{56}\text{,}
\end{displaymath}

\noindent
where $A$~--- is the atomic number of a primary particle, $d$~--- any air shower parameter that is sensitive to CR mass composition obtained in experiment (exp) or in calculation for primary protons ($p$) and iron nuclei (Fe). Here one cannot do without theoretical notion of EAS development. In work~\cite{bib:8} LDFs of responses were calculated for surface-based and underground scintillation detectors of Yakutsk array in air showers initiated by primary particles with $\E \ge 10^{17}$~eV. Calculations were performed within the frameworks of \qgs~\cite{bib:9}, \qgsii~\cite{bib:10}, \eposlhc~\cite{bib:11} and \sibyllold~\cite{bib:12} models with the use of \corsika~\cite{bib:13} code.

\begin{figure}[htb]
    \centering
    \includegraphics[width=0.65\textwidth]{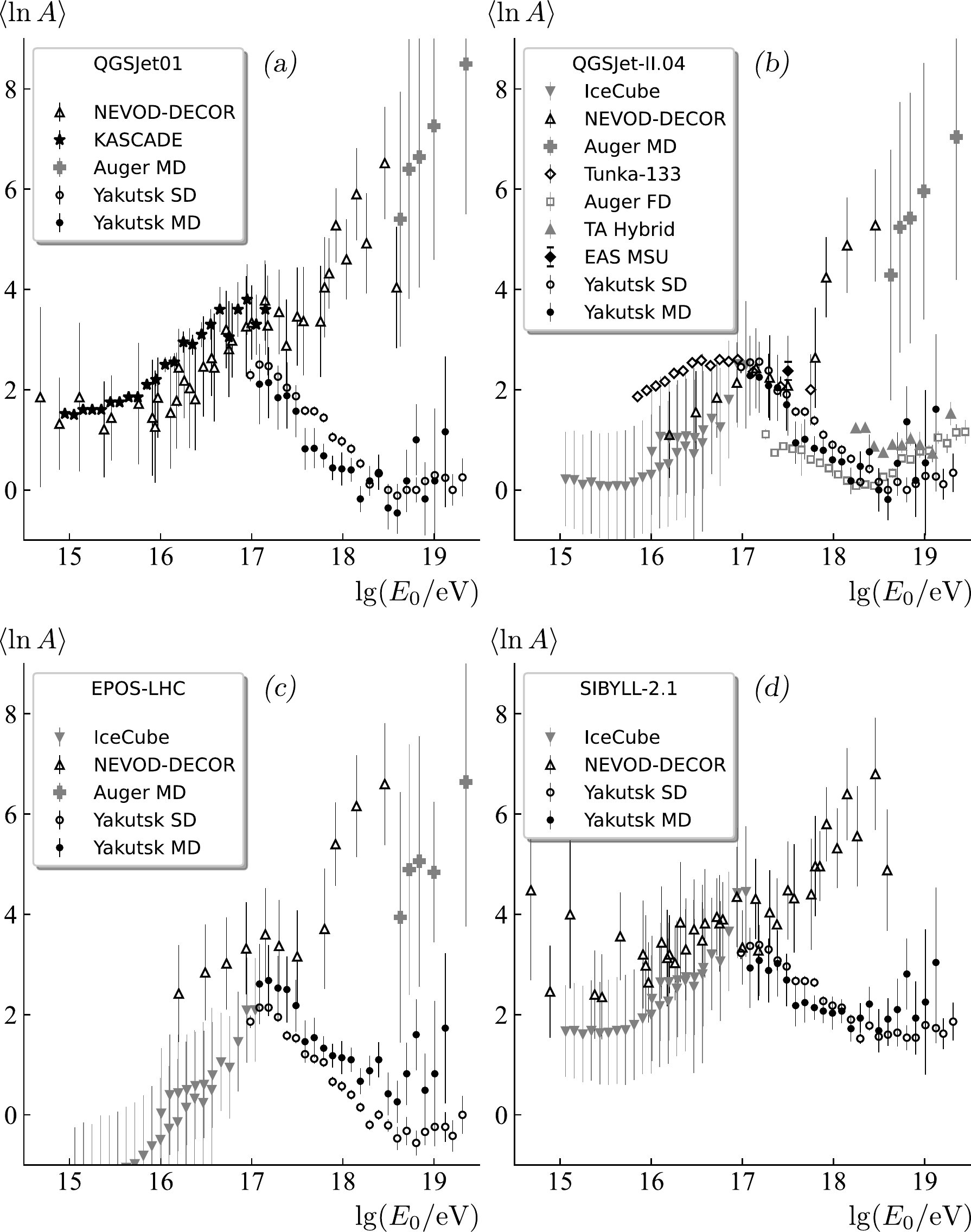}
    \caption{Energy dependencies of the CR mass compositions obtained at different EAS arrays. Empty circles~--- estimation obtained at the Yakutsk array from the data of surface detectors (SD)~\cite{bib:16, bib:17}, filled circles~--- estimation according to muon component of EAS (MD)~\cite{bib:18}. Also shown are estimations obtained from the scaling parameter $z$~\cite{bib:19} for experiments IceCube~\cite{bib:20}, NEVOD-DECOR~\cite{bib:21, bib:22}, EAS-MSU~\cite{bib:23} and The Pierre Auger Observatory (PAO)~\cite{bib:14, bib:25, bib:26}. Also shown are the data of KASCADE~\cite{bib:27} and Tunka-133~\cite{bib:28}, of fluorescent emission measurements performed at PAO (FD)~\cite{bib:29} and Telescope Array (TA)~\cite{bib:30, bib:31}}
    \label{fig:1}
\end{figure}

On \fign{fig:1} estimations of primary particles composition are shown obtained in world EAS arrays. They demonstrate an inconsistent picture. Arrays that register muons in inclined showers~--- NEVOD-DECOR and Auger~--- at $\E \ge 3 \times 10^{17}$~eV give values of $A$ which go far beyond the traditional notion. All these data have revealed the problem of ``muon excess'' in air showers~\cite{bib:14, bib:15} which calls into question all existing hadron interaction models used in high-energy physics. In the light of these facts here we are analyzing the data of Yakutsk array at $\E \simeq 10^{18}$~eV which have large statistics and high precision.

\section{Quality cuts and events processing}

\begin{table}[htb]
    \centering
    \caption{Statistics of events included in the analysis}
    \label{t:1}
    \begin{tabular}{lrrrrrr}
    \hline
    $\meancos$ & 0.95 & 0.90 & 0.85 & 0.75 & 0.65 & 0.55 \\
    \hline
    Number of showers & 2835 & 1774 & 983 & 637 & 461 & 246 \\
    \hline
    \end{tabular}
\end{table}

\begin{figure}[htb]
    \centering
    \includegraphics[width=0.65\textwidth]{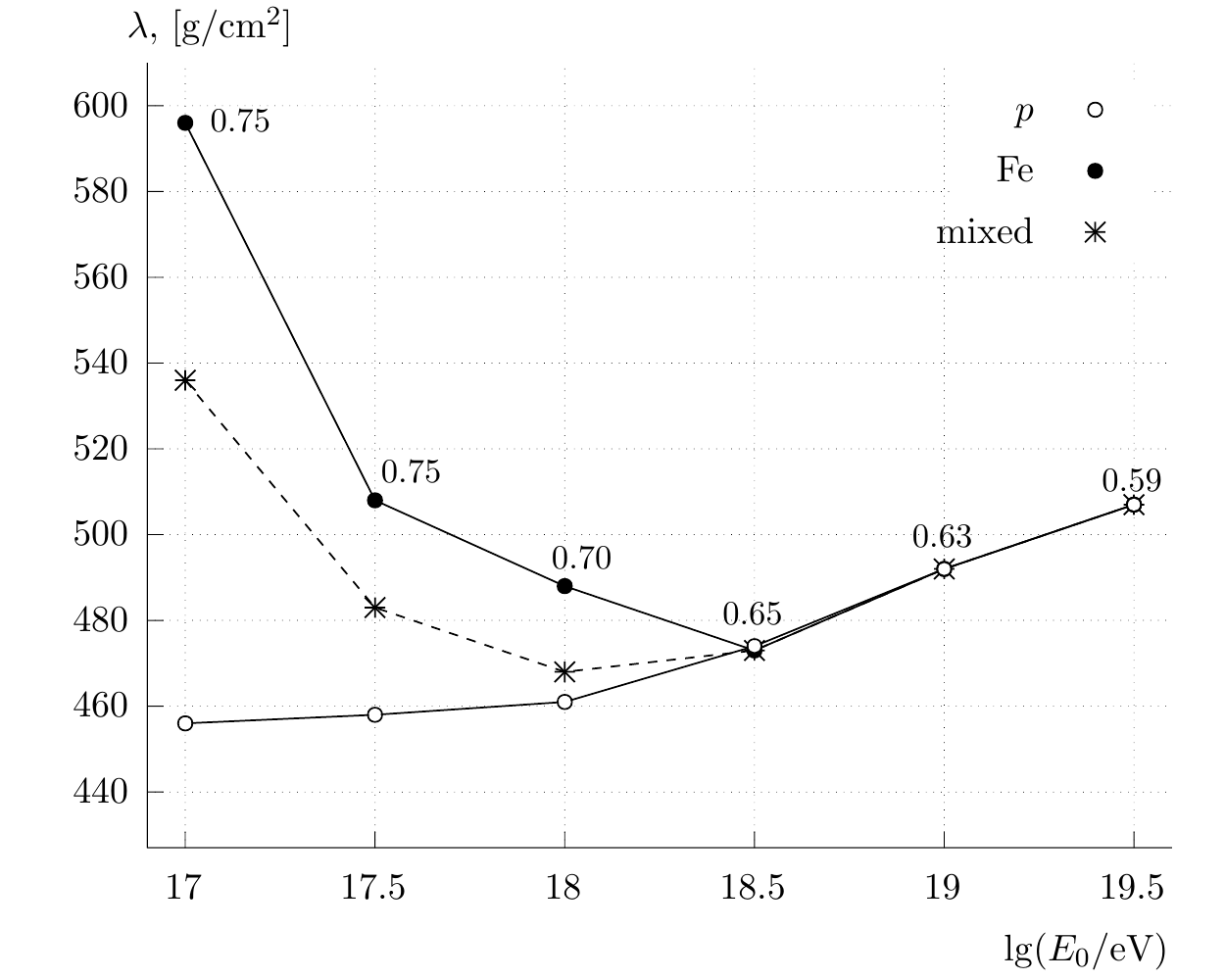}
    \caption{Energy dependence of the absorption length $\lambda$ in equation \eq{eq:3}{} used in recalculation from $\rhosSOOT$ to $\rhosSOOV$ according to \qgs{} model for primary protons ($p$), mixed composition (``mixed'') and primary iron nuclei (Fe). Numbers near data points denote the maximum allowable values of $\cos\theta$.}
    \label{fig:2}
\end{figure}

\begin{figure}[htb]
    \centering
    \includegraphics[width=0.65\textwidth]{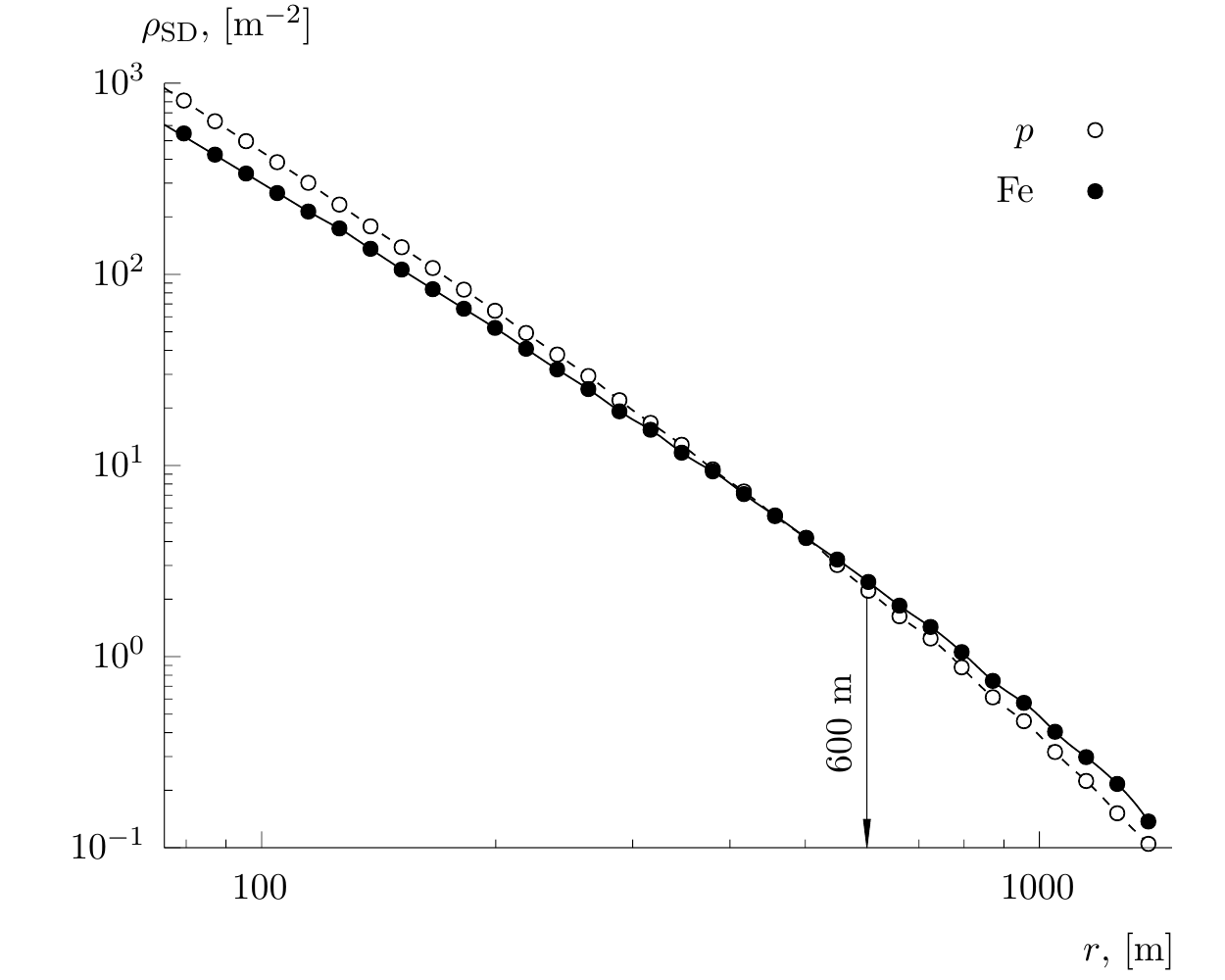}
    \caption{LDFs of charged particles in showers with primary energy $10^{18}$~eV and vertical arrival direction $\cos\theta = 0.9$ for primary protons and iron nuclei calculated within the framework of \qgsii{} model~\cite{bib:8}. The arrow denotes the distance 600~m from shower axis.}
    \label{fig:3}
\end{figure}

Here we analyze mean densities of all EAS particles $\left<\rhosRT\right>$ and muons $\left<\rhomRT\right>$ with threshold energy $\ethr \simeq 1.0 \times \sec\theta$~GeV measured with surface-based and underground detectors at $r = 300, 600$ and $1000$~m from the axis in showers with mean zenith directions $\meancos = 0.95, 0.90, 0.85, 0.75, 0.65$ and 0.55. The statistics of used events is presented in Table~\ref{t:1}. Experimental LDFs of both components were constructed within intervals $\Delta\cos\theta = 0.1$ with logarithmic step along the energy scale $\Delta\lg(\E/\text{eV}) = 0.2$. The following quality cuts were imposed on the dataset: zenith angles $\theta \le 60\degr$; axes located within a 1-km circle around the array center and reconstructed with errors below 50~m. Primary energy was estimated from relations:

\begin{gather}
    \E = (3.76 \pm 0.30)
    \times
    10^{17}
    \cdot
    \rhosSOOV^{1.02 \pm 0.02}~\text{[eV],}
    \label{eq:1} \\
    \rhosSOOV = \rhosSOOT
    \cdot
    \exp{
        \frac{
            \sec\theta - 1
        }{
            \lambda
        } \times h
    }~\text{[m}^{-2}\text{],}
    \label{eq:2}
\end{gather}

\noindent
with absorption length $\lambda$ shown of \fign{fig:2}; $h = 1020$~\depth~--- atmospheric depth of the Yakutsk array. The precision of $\rhosSOOT$ measurement in individual events was not worse than 10\%. The relation \eqn{eq:1} unambiguously connects $\E$ with $\rhosSOOV$ at any CR composition since lateral distributions of SD-detected particles in showers originated from different CR specimen intersect at axis distance $\nthick\simeq 600$~m. It is demonstrated on \fign{fig:3} where two LDFs are shown for showers with $\E = 10^{18}$~eV and $\cos\theta = 0.9$, initiated by primary protons and iron nuclei, simulated within the framework of \qgsii{} hadron interaction models.

Axes coordinates and energy estimator $\rhosRT$ were reconstructed by the modified Linsley function~\cite{bib:32}:

\begin{equation}
        \fs(r, \theta) = \rhosSOOT \cdot
    \left(
        \frac{600}{r}
    \right)^{\alpha} \cdot
    \left(
        \frac{\rM + 600}{\rM + r}
    \right)^{\bs - \alpha}\text{,}
    \label{eq:3}
\end{equation}

\noindent
where $\alpha = 1$, $\rM$~--- Moliere radius. The latter depends on air temperature $t\degr$C and pressure $P$~[mbar]:

\begin{equation}
    \rM =
    \frac{7.5 \times 10^{4}}{P}
    \times
    \frac{t}{273}~\text{.}
    \label{eq:4}
\end{equation}

\noindent
The value of $\rM$ was determined in every event (for Yakutsk seasonal average values are $\left<t\right> \simeq -18\degr$C, $\rM \simeq 70$~m). The $\bs$ parameter in equation \eq{eq:3} was determined earlier~\cite{bib:33}:

\begin{equation}
    \bs = 1.38 + 2.16 \cos\theta + 0.15
    \times
    \lg{\rhosSOOT}\text{.}
    \label{eq:5}
\end{equation}

During LDFs construction particle densities in individual showers were multiplied by the normalization ratio $\left<\E\right> / \E$, where $\left<\E\right>$ is the mean energy within a group. The normalized densities were averaged within bins $\Delta\lg(r/\text{m}) = 0.04$ along the axis distance. Average particle densities were determined with a formula

\begin{equation}
    \left<\rhos(r_i)\right> = 
    \frac{1}{N}
    \sum_{k = 1}^{N}\rho_k(r_i)\text{,}
    \label{eq:6}
\end{equation}

\noindent
where $N$ is the number of detector readings within an axis distance interval $\lg{r_i}, \lg{r_i} + 0.04$. The resulting LDFs were approximated with a function

\begin{equation}
    \rhosRT = \fsh(r,\theta) \cdot
    \left(
        \frac{600 + r_1}{r + r_1}
    \right)^g~\text{,}
    \label{eq:7}
\end{equation}

\noindent
where $a = 2$, $\rM = r_0 = 8$~m, $r_1 = 10^4$~m and $g = 10$. The second term in equation \eq{eq:7} was introduced for correction of the LDF's steepness at large distances from shower axis. In this function a transition was made from Moliere radius $\rM$ to formal parameter $r_0$ which, together with other parameters in \eq{eq:7}, provides its best agreement with average densities calculated with \eq{eq:6} on a wide range of axis distances. The $\rhosSOOT$ and $\bs$ were free parameters and were determined during a $\chi^2$ minimization procedure. The final energy values were calculated from the resulting MLDFs with the use of refined calorimetric method (see~\cite{bib:34}).

LDFs of muon component were constructed in a similar way. Average particle densities were calculated according to a formula:

\begin{equation}
    \left<\rhom(r_i)\right> = 
    \frac{1}{N_1 + N_0}
    \sum_{n = 1}^{N_1}\rho_n(r_i)\text{,}
    \label{eq:8}
\end{equation}

\noindent
where $N_1$ and $N_0$ are numbers of non-zero and zero readings of muon detectors located within a range of axis distance $(\lg{r_i}, \lg{r_i} + 0.04)$. With ``zero readings'' we denote cases when detectors haven't register a single muon while being in accepting mode. The LDFs were approximated with a function

\begin{equation}
    \rhomRT = \fmu(r,\theta)
    \cdot
    \left(
        \frac{600 + r_1}{r + r_1}
    \right)^g\text{,}
    \label{eq:9}
\end{equation}

\noindent
where $r_1 = 2000$~m, $g = 6.5$ and $\fmu(r,\theta)$ is a function proposed by Greisen~\cite{bib:35}:

\begin{equation}
    \fmu(r,\theta) = \rhomSOOT
    \cdot
    \left(
        \frac{600}{r}
    \right)^{0.75}
    \cdot
    \left(
        \frac{r_0 + 600}{r_0 + r}
    \right)^{\bmu - 0.75}\text{.}
    \label{eq:10}
\end{equation}

\noindent
Here $r_0 = 280$~m and free parameters $\bmu$ and $\rhomSOOT$ were determined during a $\chi^2$-minimization.

\section{Results and discussion}

Lateral distributions of EAS particles at Yakutsk array are measured in units of energy deposited in a plastic scintillator by vertical relativistic muons. The density of plastic scintillators used in detectors is 1.06~\dens{} and thickness is 5~cm. This energy ($E_1 = 5 \times 1.06 \times 2.217~\text{MeV} = 11.75$~MeV) is spent on ionization of scintillator's medium and is re-emitted as a flash of light with number of photons proportional to the number of particles passed through a detector (electrons, muons and high-energy gamma-photons). This flash of light is subsequently converted into electric charge (the response) by a photomultiplier tube. In real experiment the total energy deposit from all these particles $\Delta{}E_s(r)$ is measured as a conventional density in units of $\rhos(r) = \Delta{}E_s(r) / E_1$~[m$^{-2}$]. Photonic shower component is registered via pairs production and generation of $\delta$-electrons from Compton scattering. It adds a significant contribution into LDF of surface detectors.

\begin{figure}[htb]
    \centering
    \includegraphics[width=0.85\textwidth]{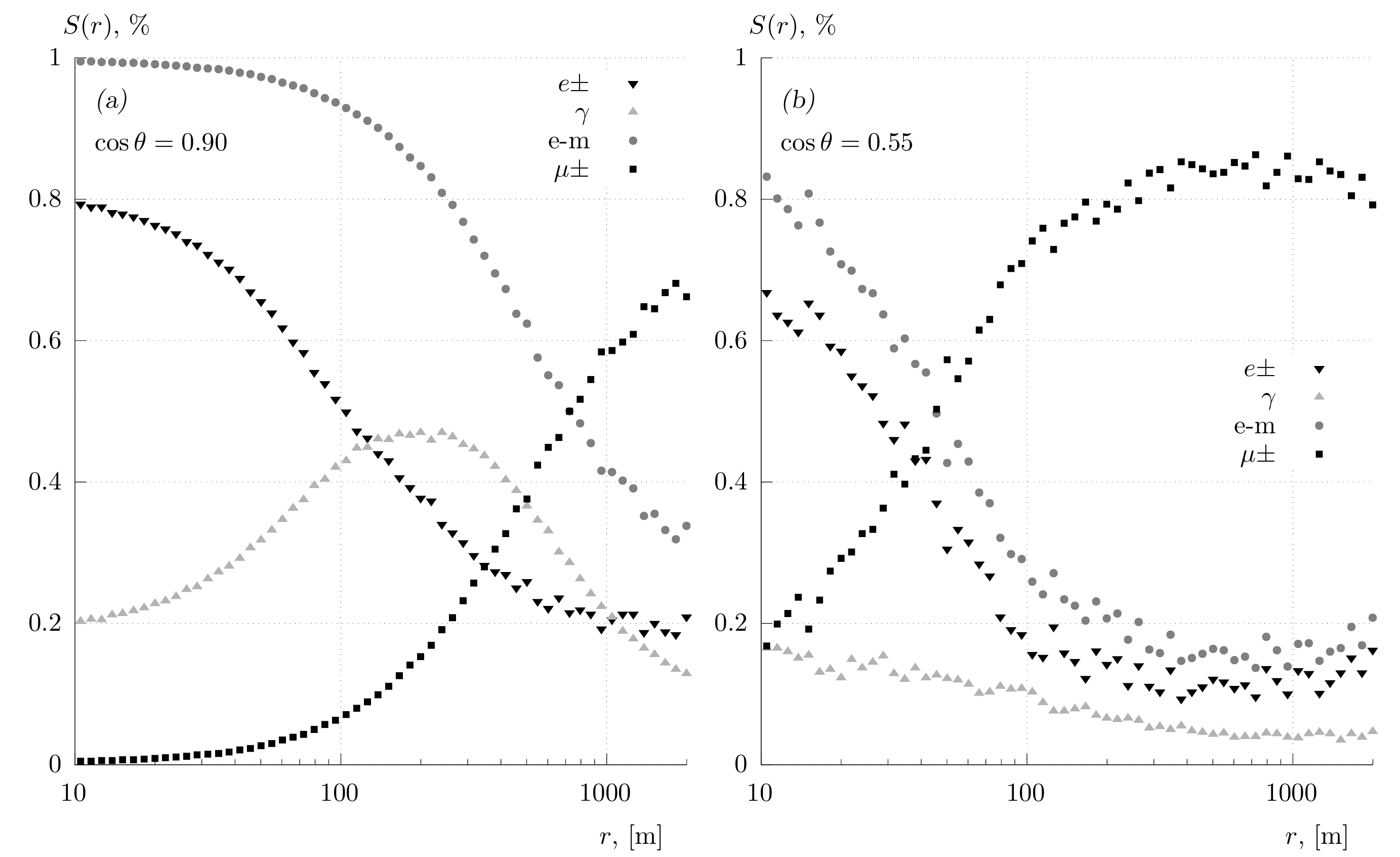}
    \caption{Relative contributions of EAS components into total SD response at $30-2000$~m from shower axis with vertical arrival direction $\cos\theta = 0.90$ {\sl (a)} and $\cos\theta = 0.55$ {\sl (b)} obtained within the framework of \qgs{} model for primary protons~\cite{bib:8}.}
    \label{fig:4}
\end{figure}

On \fign{fig:4} relative contributions $S(r)$ from all these particles in total response are shown at $r = 30 - 2000$~m from the axis in showers with zenith angles $\theta = 25.8\degr$ and $56.6\degr$ calculated within the framework of \qgs{} model for primary protons~\cite{bib:8}. For treatment of low-energy interactions \fluka{} code~\cite{bib:36} was used. Responses from particles reaching the ground level were calculated as $U_k(E, \theta)$, where $k$ is either electron, muon or gamma-photon, $E$~--- particle energy and $\theta$~--- its incident zenith angle. Calculation were performed with the account of all processes with corresponding cross-sections occurring during energy emission or absorption in the detector's shielding and inside a scintiilator. Air shower development was simulated with the use of \corsika{} code. For every set of primary parameters (primary particle mass, energy and zenith angle) 200 showers were simulated. Thin-sampling mechanism was activated in order to speed-up the simulation, with following parameters: $E_i/E_0 = 10^{-5}$, $\wmax = 10^{4}$. During the conversion into estimated particle density, total number of particles arriving at detector of given area was taken into account. Showers were averaged over the set. Particle data of simulation were used to calculate differential energy spectra in each radial bin $\Delta\lg(r/\text{m})$ as $d_k(E, r, \theta)$. The total response was defined as a sum over partial responses from all components:

\begin{equation}
    S(r) = \sum_{k = 1}^{3} \sum_{i = 1}^{I_k}
        U_k(E_i, \theta_i)
        \cdot
        d_k(E_i, r, \theta_i)\text{,}
    \label{eq:11}
\end{equation}

\noindent
where $I_k$ is the number of particle records of type $k$ hitting a detector. The density of pure electromagnetic component of EAS (e-m) on \fign{fig:4} is a sum of electrons and photons. Combined with muons ($\mu\pm$) it forms the total response $S(r)$. In inclined events at $r \ge 300$~m the muon component constitutes up to 80\% of total response in surface scintillation detectors.

\begin{figure}[htb]
    \centering
    \includegraphics[width=0.65\textwidth]{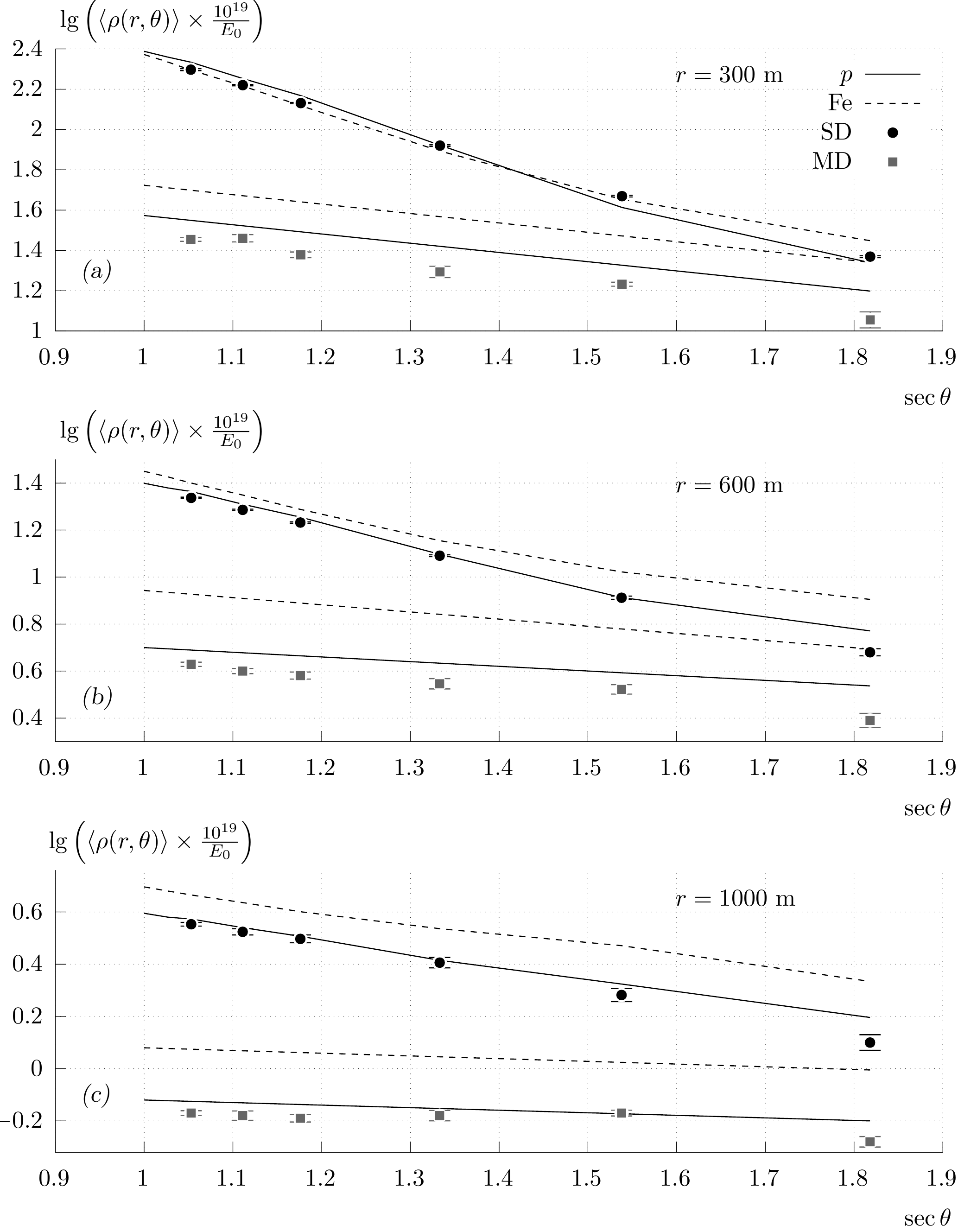}
    \caption{Zenith-angular dependencies of surface-based scintillation detectors response (SD) and underground muon detectors response (MD) on Yakutsk array at 300~m {\sl (a)}, 600~m {\sl (b)} and 1000~m {\sl (c)} from shower axis at energy $10^{18}$~eV, normalized to experimentally evaluated primary energy \eq{eq:1}. On the panel {\sl (a)} symbols denote the experimental values, lines~--- simulation results obtained within the framework of \qgs{} model for primary protons and iron nuclei. Designations on panels {\sl (b)} and {\sl (c)} are the same.}
    \label{fig:5}
\end{figure}

Further we consider values $\left<\rhos(r,\theta)\right>$ and $\left<\rhom(r, \theta)\right>$ determined with the use of equations \eq{eq:7} and \eq{eq:10}. We present summary errors that include both statistical and systematical constituents. In practice it is hard and not always practical to distinguish one from another. On \fign{fig:5} are shown zenith-angular dependencies of particle densities $\lg((\left<\rhosRT\right> / \E) \times 10^{19})$ and $\lg((\left<\rhomRT\right> / \E) \times 10^{19})$ normalized by primary energy estimated according to \eq{eq:1}. Experimental values are compared to the predictions of \qgs{} model for primary protons and iron nuclei. It is seen that densities of both EAS components measured in different zenith-angular intervals do not contradict the hypothesis of pure proton CR composition at energy $\nthick\sim 10^{18}$~eV. Muons in experimental data demonstrate a $\nthick\simeq 10$\% deficit when compared to simulation results.

\begin{figure}[htb]
    \centering
    \includegraphics[width=0.65\textwidth]{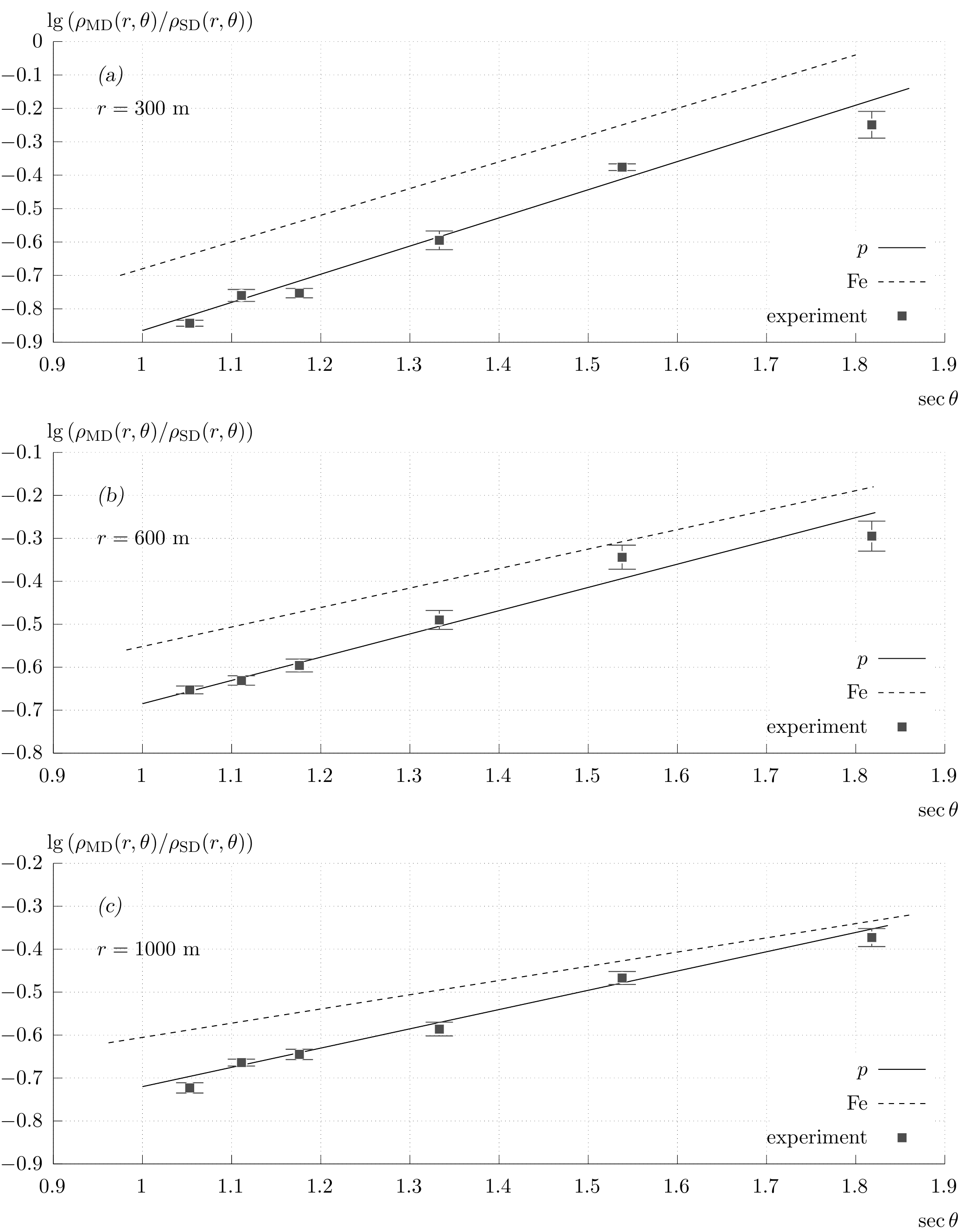}
    \caption{Zenith-angular dependencies of the muon fraction ($1.0 \times \sec\theta$-GeV threshold) relative to the total number of particles registered by surface-based detectors at 300~m {\sl (a)}, 600~m {\sl (b)} and 1000~m {\sl (c)} from shower axis at primary energy $10^{18}$~eV. Lines~--- results of calculations within the framework of \qgs{} model{} for primary protons ($p$) and iron nuclei (Fe).}
    \label{fig:6}
\end{figure}

Zenith-angular dependencies of the muon fraction $\lg(\left<\rhomRT\right> / \left<\rhosRT\right>)$ in total SD response is shown on \fign{fig:6}. These values are directly connected with the composition of primary CR particles. It is seen that the best agreement between theory and experiment is observed in the case of \qgs{} model. From this figure one can assume that energy reconstructed for all the data presented on \fign{fig:5} according to equation \eq{eq:1} is probably overestimated by factor 1.1. This hypothesis has yet the right to exist but needs further comprehensive verification. Depths of maximum of cascade curves in showers with energy $10^{18}$~eV initiated by primary protons and iron nuclei obtained within the frameworks of \qgs{} model are equal to $721 \pm 30$ and $636 \pm 20$~\depth{} accordingly. In the last angular interval ($\sec\theta = 1.818$) the experimental densities are lower by 20\% compared to simulation results. This can be explained by the fact that primary energy calculated according to equation \eq{eq:1} turned out to be higher than values in first five intervals by factor 1.1. This tendency can probably be more pronounced in more inclined EASs. It is seen from \fign{fig:6} that at distances $r \simeq 300$~m from the axis in showers with zenith angles $\nthick\simeq 60\degr$, the fraction of muons can be close to 1. This is a hint that surface-based and underground detectors register virtually the same component os air showers, namely~--- muons with energy $\nthick\ge 2$~GeV. The results of NEVOD-DECOR and Auger experiments shown on \fign{fig:1}, which both agree with abnormally heavy CR composition at energy $\nthick\sim 10^{18}$~eV, can be explained within the framework of this assumption. They register muons in strongly inclined showers ($\theta > 60\degr$), where correct measurement of muon densities and estimation of primary energy are complex and important tasks.

\section{Conclusions}

The longstanding measurements of the spatial structure of EAS particles distribution performed with the use of surface-based and underground detectors at the Yakutsk array and their comparison with LDFs obtained in theoretical calculations, have again allowed us to critically evaluate the mass composition of CR in energy range $\nthick\simeq 10^{18}$~eV, where a significant body of experimental material have been accumulated. The combined analysis of zenith-angular dependencies for two air shower components measured with similar scintillation detectors, equally calibrated by the frequency of cosmic muon background, have demonstrated  satisfactory agreement with the predictions of \qgs{} model in the case of proton composition of primary particles with the considered energy. This conclusion is based on the measurements of $6 \times 3 = 18$ independent density values $\left<\rhosRT\right>$ and the same number of independent values of $\left<\rhomRT\right>$. In the latter case the registration threshold in inclined showers increases from 1.0 to 1.8~GeV. In both simulation and experiment the detector responses were calculated~--- the conditional number of particles measured in units of vertical relativistic muons in the detector placed under a layer of lead with 35-cm thickness. Here we have omitted considering the \qgsii{} model since its predictions do not contradict the above-mentioned conclusion on CR mass composition~\cite{bib:16, bib:17, bib:18}. Our plans are to continue this study at other energies of EAS. Also we will consider in detail predictions of \eposlhc{} and \sibyllold{} models whose predictions disagree with the data of Yakutsk array, especially the latter one~\cite{bib:16, bib:17, bib:18}.


\begin{thebibliography}{37}

\bibitem{Grieder(2010)}
    P.~K.~F.~Grieder, \DOI{10.1007/978-3-540-76941-5}{\sl Extensive Air Showers: High Energy Phenomena and Astrophysical Aspects}{} (Springer, Berlin, 2010).

\bibitem{bib:2}
    A.~V.~Glushkov, V.~M.~Grigoriev, N.~N.~Efimov, M.~I.~Pravdin, O.~S.~Diminstein, and V.~P.~Sokurov, in \href{https://ui.adsabs.harvard.edu/abs/1979ICRC....8..158G}{{\sl Proc. of the 16th ICRC, Kyoto}, Vol.~8, edited by S.~Miyake (Nihon Gakujutsu Kaigi, Tokyo, 1979) p.~158}.

\bibitem{bib:3}
A.~V.~Glushkov, {\sl Lateral distribution and total flux of Cherenkov light emission in EAS with primary energy $\E \nthick\sim 10^{17}$~eV}. PhD thesis. SINP MSU (Moscow, 1982) [in Russian].

\bibitem{bib:4}
A.~V.~Glushkov, L.~G.~Dedenko, N.~N.~Efimov, N.~N.~Efremov, I.~T.Makarov, P.~D.~Petrov and M.~I.~Pravdin. Izv. AN SSSR. Ser.: Phys. {\bf 55}, 2166 (1986) [in Russian].

\bibitem{bib:5}
A.~V.~Glushkov, M.~I.~Pravdin, I.~E.~Sleptsov, V.~R.~Sleptsova and N.~N.~Kalmykov. YaF. {\bf 63}, 1557 (2000) [in Russian].

\bibitem{bib:6}
    A.~V.~Glushkov and A.~V.~Saburov. \DOI{10.1134/S0021364013230057}{JETP Lett. {\bf 98}, 589 (2014)}.

\bibitem{bib:7}
    E.~G.~Berezhko, S.~P.~Knurenko, L.~T.~Ksenofontov. \DOI{10.1016/j.astropartphys.2012.04.014}{Astropart. Phys. {\bf 36}, 31 (2013)}.

\bibitem{bib:8}
A.~V.~Saburov, {\sl Lateral distribution particles in EAS with energy above $10^{17}$~eV according to the data of Yakutsk array.} PhD thesis. INR RAS (Moscow, 2018) [in Russian].

\bibitem{bib:9}
    N.~N.~Kalmykov, S.~S.~Ostapchenko, A.~I.~Pavlov. \DOI{10.1016/S0920-5632(96)00846-8}{Nucl. Phys. B~--- Proc. Suppl. {\bf 52}, 17 (1997)}.

\bibitem{bib:10}
    S.~Ostapchenko. \DOI{10.1103/PhysRevD.83.014018}{Phys. Rev. D. {\bf 83}, 014018 (2011)}. \arXiv{1010.1869}{hep-ph}.

\bibitem{bib:11}
    T.~Pierog, Iu.~Karpenko, J.~M.~Katzy, E.~Yatsenko, and K.~Werner. \DOI{10.1103/PhysRevC.92.034906}{Phys. Rev. C. {\bf 92}, 034906 (2015)}. \arXiv{1306.0121}{hep-ph}.

\bibitem{bib:12}
    E.-J.~Ahn, R.~Engel, T.~K.~Gaisser, P.~Lipari, and T.~Stanev. \DOI{10.1103/PhysRevD.80.094003}{Phys. Rev. D. {\bf 80}, 094003 (2009)}. \arXiv{0906.4113}{hep-ph}.

\bibitem{bib:13}
D.~Heck, J.~Knapp, J.~N.~Capdevielle, G.~Schatz and T.~Thouw, \href{https://www.iap.kit.edu/corsika/70.php}{\sl CORSIKA: A Monte Carlo Code to Simulate Extensive Air Showers}, FZKA 6019. (Forschungszentrum Karlsruhe, 1988).

\bibitem{bib:14}
    A.~Aab, P.~Abreu, M.~Aglietta, E.-J.~Ahn, I.~Al~Samarai, I.~F.~M.~Albuquerque, I.~Allekotte, J.~Allen, P.~Allison, A.~Almela, J.~Alvarez Castillo, J.~Alvarez-Mu\~niz, R.~Alves Batista, M.~Ambrosio, A.~Aminaei, L.~Anchordoqui et al. (for the Pierre Auger Collab.). \DOI{10.1103/PhysRevD.91.032003}{Phys. Rev. D. {\bf 91}, 032003 (2015)}. \arXiv{1408.1421}{astro-ph.HE}.

\bibitem{bib:15}
    F.~Gesualdi, A.~D.~Supanitsky and A.~Etchegoyen. \DOI{10.1103/PhysRevD.101.083025}{Phys. Rev. D {\bf 101}, 083025 (2020)}. \arXiv{2003.03385}{astro-ph.HE}.

\bibitem{bib:16}
    A.~V.~Glushkov, M.~I.~Pravdin, A.~V.~Saburov. \DOI{10.1134/S0021364014230052}{JETP Lett. {\bf 100}, 695 (2015)}.

\bibitem{bib:17}
    A.~V.~Glushkov, M.~I.~Pravdin, A.~V.~Saburov. \DOI{10.1134/S106377371810002X}{Astron. Lett. {\bf 44}, 588 (2018)}.

\bibitem{bib:18}
    A.~V.~Glushkov, A.~V.~Saburov. \DOI{10.1134/S0021364019090091}{JETP Lett. {\bf 109}, 559 (2019)}.

\bibitem{bib:19}
    H.~P.~Dembinski, J.~C.~Arteaga-Vel\'{a}zquez, L.~Cazon, R.~Concei\c{c}\~{a}o, J.~Gonzalez, Y.~Itow, D.~Ivanov, N.~N.~Kalmykov, I.~Karpikov, S.~M\"{u}ller, T.~Pierog, F.~Riehn, M.~Roth, T.~Sako, D.~Soldin, R.~Takeishi et al. (for EAS-MSU, IceCube, KASCADE-Grande, NEVOD-DECOR, Pierre Auger Observatory, SUGAR, Telescope Array, Yakutsk EAS Array). \DOI{10.1051/epjconf/201921002004}{EPJ Web Conf. {\bf 210}, 02004 (2019)}.

\bibitem{bib:20}
    J.~G.~Gonzalez (for the IceCube Collaboration). \DOI{10.1051/epjconf/201920803003}{EPJ Web of Conf. {\bf 208}, 03003 (2019)}.

\bibitem{bib:21}
    A.~G.~Bogdanov, D.~M.~Gromushkin, R.~P.~Kokoulin, G.~Mannocchi, A.~A.~Petrukhin, O.~Saavedra, G.~Trinchero, D.~V.~Chernov, V.~V.~Shutenko, I.~I.~Yashin. \DOI{10.1134/S1063778810110074}{Phys. Atom. Nucl. {\bf 73}, 1852 (2010)}.

\bibitem{bib:22}
    A.~G.~Bogdanov, R.~P.~Kokoulin, G.~Mannocchi, A.~A.~Petrukhin, O.~Saavedra, V.~V.~Shutenko, G.~Trinchero, I.~I.~Yashin. \DOI{10.1016/j.astropartphys.2018.01.003}{Astropart. Phys. {\bf 98}, 13 (2018)}.

\bibitem{bib:23}
    Yu.~A.~Fomin, N.~N.~Kalmykov, I.~S.~Karpikov, G.~V.~Kulikov, M.~Yu.~Kuznetsov, G.~I.~Rubtsov, V.~P.~ Sulakov, S.~V.~Troitsky. \DOI{10.1016/j.astropartphys.2017.04.001}{Astropart. Phys. {\bf 92}, 1 (2017)}. \arXiv{1609.05764}{astro-ph.HE}.

\bibitem{bib:25}
    A.~Aab, P.~Abreu, M.~Aglietta, E.-J.~Ahn, I.~Al~ Samarai, I.~F.~M.~Albuquerque, I.~Allekotte, J.~D.~Allen, P.~Allison, A.~Almela, J.~Alvarez Castillo, J.~Alvarez-Mu\~niz, R.~Alves Batista, M.~Ambrosio, A.~Aminaei, L.~Anchordoqui et al. (for the Pierre Auger Collab.). \DOI{10.1103/PhysRevLett.117.192001}{Phys. Rev. Lett. {\bf 117}, 192001 (2016)}. \arXiv{1610.08509}{hep-ex}.

\bibitem{bib:26}
    S.~M\"{u}ller (for the Pierre Auger Collaboration). \DOI{10.1051/epjconf/201921002013}{EPJ Web Conf. {\bf 210}, 02013 (2019)}. 

\bibitem{bib:27}
    H.~Ulrich, T.~Antoni, W.~D.~Apel, F.~Badea, K.~Bekk, A.~Bercuci, H.~Bl\"{u}mer, E.~Bollmann, H.~Bozdog, I.~M.~Brancus, C.~Buttner, A.~Chilingarian, K.~Daumiller, P.~Doll, J.~Engler, F.~Fessler et al. (KASCADE Collab.), in: \href{https://ui.adsabs.harvard.edu/abs/2001ICRC....1...97U/}{{\sl Proc. 27th ICRC, Hamburg}, Vol.~2, edited by K.-H. Kampert, G. Hainzelmann, C. Spiering. Copernicus (Berlin, 2001) p.~97}.

\bibitem{bib:28}
    V.~V.~Prosin,  S.~F.~Berezhnev,  N.~M.~Budnev,  A.~Chiavassa,  O.~A.~Chvalaev,  O.~A.~Gress,  A.~N.~Dyachok,  S.~N.~Epimakhov,  N.~I.~Karpov,  N.~N.~Kalmykov, E.~N.~Konstantinov, A.~V.~Korobchenko, E.~E.~Korosteleva, V.~A.~Kozhin, L.~A.~Kuzmichev, B.~K.~Lubsandorzhiev et al. \DOI{10.1016/j.nima.2013.09.018}{Nucl. Instr. Meth. A. {\bf 756}, 94 (2014)}.

\bibitem{bib:29}
    J.~Bellido for the Pierre Auger Collaboration, in: \DOI{10.22323/1.301.0506}{{\sl Proc. of the 35th ICRC, Busan}, 2017, PoS(ICRC2017)506}.

\bibitem{bib:30}
    R.~U.~Abbasi, M.~Abe, T.~Abu-Zayyad, M.~Allen, R.~Azuma, E.~Barcikowski, J.~W.~Belz, D.~R.~Bergman, S.~A.~Blake, R.~Cady, B.~G.~Cheon, J.~Chiba, M.~Chikawa, T.~Fujii, K.~Fujita, M.~Fukushima et al. \DOI{10.3847/1538-4357/aabad7}{ApJ. {\bf 858}, 76 (2018)}. \arXiv{1801.09784}{astro-ph.HE}.

\bibitem{bib:31}
    R.~U.~Abbasi, M.~Abe, T.~Abu-Zayyad, M.~Allen, R.~Azuma, E.~Barcikowski, J.~W.~Belz, D.~R.~Bergman, S.~A.~Blake, R.~Cady, B.~G.~Cheon, J.~Chiba, M.~Chikawa, T.~Fujii, K.~Fujita, M.~Fukushima et al. \DOI{10.1103/PhysRevD.99.022002}{Phys. Rev. D. {\bf 99}, 022002 (2019)}. \arXiv{1808.03680}{astro-ph.HE}.

\bibitem{bib:32}
J.~Linsley, L.~Scarsi, B.~Rossi. J. Phys. Soc. Japan. {\bf 17}, Suppl. A-III, 91 (1962).

\bibitem{bib:33}
A.~V.~Glushkov, O.~S.~Diminstein, N.~N.~Efimov, L.~I.~Kaganov and M.~I.~Pravdin, in {\sl Characteristics of extensive air showers from ultra-high energy cosmic rays.} Yakutsk, YaF SO AN SSSR, p.~45 (1976) [in Russian].

\bibitem{bib:34}
    A.~V.~Glushkov, M.~I.~Pravdin, \& A.~V.~Saburov. \DOI{10.1134/S106377881804004X}{Phys. Atom. Nucl. {\bf 81}, 535 (2018)}.

\bibitem{bib:35}
    K.~Greisen. \DOI{10.1146/annurev.ns.10.120160.000431}{Ann. Rev. Nucl. Sci. {\bf 10}, 63 (1960)}.

\bibitem{bib:36}
    A.~Ferrari, P.~R.~Sala, A.~Fass\`{o}, J.~Ranft. \href{https://doi.org/10.5170/CERN-2005-010}{\sl FLUKA : A multi-particle transport code (program version 2005)}, CERN-2005-010. (CERN, Geneva, 2005). 

\bibitem{bib:38}
    F.~Riehn, R.~Engel, A.~Fedynitch, T.~K.~Gaisser and T.~Stanev. \DOI{10.1103/PhysRevD.102.063002}{Phys. Rev. D. {\bf 102}, 063002 (2020)}. \arXiv{1912.03300}{hep-ph}.

\end{thebibliography}
\end{document}